\documentclass[final,5p,times,twocolumn,colorlinks=true,urlcolor=blue,linkcolor=blue,citecolor=blue,sort&compress]{elsarticle} 
\journal{Physics Letters B}

\usepackage{graphicx}
\usepackage{amsmath}
\usepackage{amssymb}
\usepackage{orcidlink} 
\usepackage{siunitx}  

\begin{document}
\begin{frontmatter}
\title{Extraction of $\sigma_{TT}$ for Proton, Neutron, Deuteron and $^3$He from Quasi-real Photon Scattering}
%
%
\author[UVa]{YiLei~Li~\orcidlink{0000-0001-7901-9860}}
\author[UVa]{B.~Callahan}  
\author[JLab]{M.M.~Dalton\corref{cor1}~\orcidlink{0000-0001-9204-7559}} 
\author[JLab]{A.~Deur~\orcidlink{0000-0002-2203-7723}}
\author[UVa]{O.~Larson}
\author[UVa]{A.~Rask~\orcidlink{0009-0004-1264-3676}}
\author[ODU]{D.W.~Upton~\orcidlink{0000-0002-1456-7572}}
\author[UVa]{X.~Zheng~\orcidlink{0000-0001-7300-2929}}
\affiliation[UVa]{
  organization={University of Virginia},
  city={Charlottesville},
  postcode={22904},
  state={VA},
  country={USA}
}
\affiliation[JLab]{
  organization={Thomas Jefferson National Accelerator Facility},
  city={Newport News},
  postcode={23606},
  state={VA},
  country={USA}
}
\affiliation[ODU]{
  organization={Old Dominion University},
  city={Norfolk},
  postcode={23529},
  state={VA},
  country={USA}
}
\cortext[cor1]{Corresponding author: dalton@jlab.org} 
\begin{abstract}
We report on an extraction of the polarized photoproduction cross-section for the proton, deuteron, neutron and $^3$He, obtained by extrapolating electron scattering data to the real photon point. 
The data are from the Jefferson Lab E97-110 ($^3$He) and CLAS EG4 (proton and deuteron) experiments. Information on the neutron is extracted from the deuteron or $^3$He data using the weak binding approximation. 
Comparing with data obtained with real photons, we find that while the proton results agree, our results on the deuteron and the neutron exhibit a larger strength in the $\Delta(1232)$ region, and are more consistent with isospin symmetry when compared with the proton results.
\end{abstract}
\end{frontmatter}

\section{Introduction}

Ever since it was observed that quarks and gluons form the proton spin in a non-trivial manner~\cite{EuropeanMuon:1987isl}, the study of the spin structure of the nucleon has been an active field for both experimental and theoretical research~\cite{Anselmino:1994gn,Deur:2018roz,Ji:2020ena}. 
Crucial ingredients of these studies are spin sum rules, namely relations connecting integrals over dynamical quantities to global properties of the studied particle such as mass or spin. The integrated observables are cross sections or structure functions. 
A prominent sum rule is that of Gerasimov, Drell and Hearn (GDH)~\cite{Gerasimov:1965et,Drell:1966jv}:
\begin{equation}
I_{\rm GDH} \equiv \int_{\nu_0}^{\infty}\frac{\sigma_{\rm TT}(\nu)}{\nu}d\nu=-\frac{4 \pi^2 S \alpha \kappa^2}{M^2},
\label{eq:gdh}
\end{equation}
where $\sigma_{\rm TT}(\nu) \equiv \sigma_A(\nu)-\sigma_P(\nu)$, with $\sigma_A$ and $\sigma_P$ the photoproduction  cross-sections for cases where the photon spin is antiparallel and parallel to the target spin, respectively. Here,
$\nu$ is the photon energy,
$\nu_0$ the photoproduction threshold, 
$M$ the mass of the target particle, 
$S$ its spin, 
$\kappa$ its anomalous magnetic moment and 
$\alpha$ the electromagnetic coupling.

The GDH integrand $\sigma_{\rm TT}$ itself informs us of how different reactions contribute to $I_{\rm GDH}$. For example, if the target is a nucleon, most of the integral strength would come from the photoproduction of the $\Delta(1232)$ resonance. While if the target is a lepton, with $\kappa=0$, the positive and negative parts of $\sigma_{\rm TT}$ should cancel exactly to fulfill the sum rule~\cite{Altarelli:1972nc}. 
Additionally, one expects that only single quarks participate in the reaction at large $\nu$, and any deviation in $\sigma_{\rm TT}$ from such behavior could indicate new phenomena such as possible substructure of quarks~\cite{Dalton:2020wdv}. 
{More generally, as a polarized cross-section $\sigma_{\rm TT}$ contains crucial information on the nucleon spin structure, its spectrum helps isolates overlapping nucleon resonances, and it provides the only means known to access the forward spin polarizability $\gamma_0$~\cite{Deur:2018roz}.}
There are two extensive data sets covering a large range of $\nu$ values  for $\sigma_{\rm TT}$ of real photons. One was measured by the {\it GDH collaboration} on the proton~\cite{
GDH:2001zzk,GDH:2003xhc,Dutz:2004zz} and deuteron~\cite{Ahrens:2006yx, Ahrens:2009zz} using the MAMI and ELSA (M-E) accelerators, see review~\cite{Helbing:2006zp}. 
The other was obtained by the {\it A2 collaboration} for $^3$He in the $\Delta(1232)$ region~\cite{AguarBartolome:2013mga} and more recently for the proton and the deuteron~\cite{Pedroni:2026eqj}. Information on the neutron was extracted from both data sets~\cite{Pedroni:2026eqj, GDH:2005noz}. 
Several experiments complemented the $^3$He  data at low $\nu$~\cite{Laskaris:2013ehq,Laskaris:2015wma,Laskaris:2020ddr}, and one is planned for high $\nu$~\cite{Dalton:2020wdv}. 

With the GDH sum rule generalized to virtual photons~\cite{Anselmino:1988hn, Ji:1999mr,  Drechsel:2000ct, Drechsel:2004ki}, a substantial body of electroproduction data is now available for $\sigma_{\rm TT}(\nu, Q^2)$, where $Q^2$ denotes the absolute value of the four-momentum transfer squared between the lepton and the target, {\it i.e.} the virtuality of the exchanged photon in the Born approximation~\cite{HERMES:2000apm,HERMES:2002gmr,HERMES:1998pau,Amarian:2002ar,CLAS:2008xos,CLAS:2015otq,CLAS:2017qga,E97-110:2021mxm,JeffersonLabE97-110:2019fsc,CLAS:2021apd,CLAS:2017ozc,CLAS:2024fcf,Deur:2021klh}. 
This generalized GDH integrand can be written as 
\begin{equation}
\sigma_{TT}(\nu)=\frac{4\pi^2\alpha}{M\nu}~A_1F_1(\nu)~,
\label{eq:sigtt-A1F1}
\end{equation}
where $A_1$ is the virtual photon asymmetry and $F_1$ is the unpolarized structure function. 
It can also be related to the polarized structure function $g_1$, $g_2$ as 
\begin{equation}\sigma_{TT}(\nu, Q^2) = \frac{4\pi^2\alpha}{MK_{\gamma}}\left[g_1(\nu,Q^2)-\gamma^2g_2(\nu,Q^2)\right]~,\label{eq:sigTT_g1g2}
\end{equation}
where $\gamma^2 = Q^2/\nu^2$
and $K_{\gamma}$ is the virtual photon equivalent energy~\cite{CLAS:2024fcf}.
Some of the electroproduction data reached $Q^2$ values as low as 0.01~GeV$^2$, i.e. ``quasi-real'' photons. These experiments were designed to test Eq.~(\ref{eq:gdh}) for the proton~\cite{CLAS:2021apd}, deuteron~\cite{CLAS:2017ozc}, and neutron~\cite{CLAS:2024fcf} through extrapolation of the generalized GDH integral $I_{\rm GDH}(Q^2) \equiv \int_{\nu_0}^{\infty} \sigma_{\rm TT}(\nu,Q^2)/\nu d\nu $ to the real photon limit $Q^2\rightarrow 0$, as well as to test predictions from chiral perturbation theory~\cite{Bernard:1995dp}.   
Overall, the electroproduction data {do not challenge the GDH expectations.}

The ``quasi-real'' photon data~\cite{E97-110:2021mxm,JeffersonLabE97-110:2019fsc,CLAS:2021apd,CLAS:2017ozc,CLAS:2024fcf,Deur:2021klh} offer an opportunity to confirm, statistically improve, and complement the existing photoproduction measurements. 
Furthermore, comparing the two data sets will test the strategy of performing electroproduction measurements at very low $Q^2$ and extrapolating them to the real-photon limit $Q^2 = 0$, in order to study more photoproduction observables. This method and direct photoproduction are complementary since the techniques are quite different: with the photon beam, all the reaction final states must in principle be detected  (exclusive reaction) while with the lepton beam, detecting the scattered lepton suffices (inclusive reaction). 
Finally, the extensive $^3$He electroproduction data can be used to extract information on the neutron and to compare with that extracted from the proton and deuteron. 
In particular, it provides a test whether the weak-binding approximation (WBA) of light nuclei~\cite{Kahn:2008nq,Ethier:2013hna,Ethier:2014bua,Tropiano:2018quk} is applicable to $Q^2=0$, see Sect.~\ref{sec:WBA}.

In the following, we report on extrapolating to the real photon point the Jefferson Lab (JLab) electron scattering data on $\sigma_{\rm TT}(\nu, Q^2)$ for the proton and deuteron from the EG4 experiment~\cite{CLAS:2024fcf}, and for $^3$He from the E97-110 experiment~\cite{E97-110:2021mxm}. 

\section{Data description and extrapolation for $p$, $d$, and $^3$He}\label{sec:pdhe}

The EG4 Experiment took place in Hall B of JLab using the CLAS detector~\cite{CLAS:2003umf} and polarized NH$_3$ and ND$_3$ targets.  
Results from EG4 include structure functions $A_1F_1$ and $g_1$ for the proton and deuteron, and for the neutron extracted using WBA.
Experiment E97-110 occurred in Hall A of JLab~\cite{Alcorn:2004sb} using the High Resolution Spectrometers and a polarized $^3$He target. 
Results from E97-110 include $\sigma_{TT}$, and structure functions $g_1$ and $g_2$ for $^3$He. 
Specific experimental and analysis details for the experiments are in Refs.~\cite{JeffersonLabE97-110:2019fsc,CLAS:2024fcf}.

We extrapolated the EG4 and E97-110 data to the real photon point $Q^2=0$ by fitting them in the region $Q^2<0.2$~GeV$^2$. 
For EG4, we extrapolated $A_1F_1$ of the proton and the deuteron to $Q^2=0$ and then formed $\sigma_{TT}$ using Eq.~(\ref{eq:sigtt-A1F1}). 
For $^3$He data from E97-110, we extrapolated $\sigma_{TT}(\nu, Q^2)$ to $Q^2=0$ directly. 
The primary fitting function is linear, which was found to describe the data well and produced reduced $\chi^2$ distributed uniformly in $\nu$ and with average values of 0.7, 0.9 and 0.8 for the proton, deuteron, and $^3$He, respectively. 
{The uncertainty of the extrapolation was determined by performing extrapolation fits for all combinations of dropped points (excluding up to half of the dataset) and perturbing the data points within their corresponding uncertainties.
The uncertainty was the standard deviation of a normal function fit to the resulting set of extrapolated values.}

\begin{figure*}[!htbp]
\includegraphics[width=0.99\textwidth]{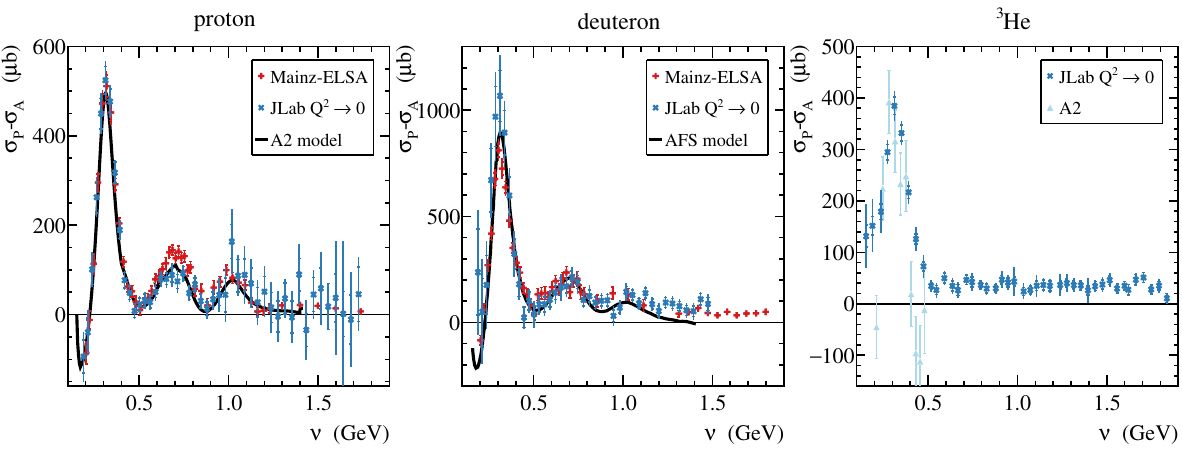}
\caption{
$\sigma_{TT}$ for the proton (left), deuteron (center) and $^3$He (right) vs. $\nu$, for photoproduction data (red $+$) and for electroproduction data extrapolated to $Q^2=0$ (blue $\times$).
For the extrapolated electroproduction data, the inner error bars are from the statistical uncertainty of the original data and the outer error bars are the total uncertainty with the statistical and the extrapolation uncertainties added in quadrature.
The peak at $\nu=0.34$~GeV in each panel signals the $\Delta(1232)$. The second and third resonance regions are clearly visible for the proton and are partly smeared by Fermi motion for the deuteron and $^3$He cases. The black curves show the models, see texts for details. }
\label{fig:sigtt_pd3He}
\end{figure*}

Our results for the proton, deuteron, and $^3$He are shown in Fig.~\ref{fig:sigtt_pd3He} along with the photoproduction data, and model predictions.  
For the proton, the model from Ref.~\cite{Pedroni:2026eqj} is a sum of final states $N\pi$~\cite{Strakovsky:2022tvu}, $N\pi\pi$~\cite{Fix:2005if}, $N\eta$~\cite{Tiator:2018heh}, and $K\Lambda+K\Sigma$~\cite{Matveev:2019igl,Sarantsev:2019xxm}. 
For the deuteron, the model from  Arenh\"{o}vel, Fix and Schwamb (AFS)~\cite{Arenhovel:2004ha} includes $NN\pi$, $NN\pi\pi$, $NN\eta$, $d\pi^0$ coherent pion production, and $pn$ photodisintegration.
{It can be noted that the two deuteron data points at the lowest $\nu$ disagree with the model, possibly due to difficulties in controlling the background from quasi-elastic scattering. This may also explain the disagreement between the $^3$He electroproduction and photoproduction data in the lowest $\nu$ bins.}

We quantify the comparison between the photo- and electroproduction data sets by integrating the spectra over $\nu$ ranges $[0.22,0.5]$, $[0.5,0.88]$ and $[0.88,1.18]$ GeV, which approximately correspond to the $\Delta(1232)$, and the second and third resonance regions, respectively, see Table~\ref{tab:integ}.
The proton results agree with the photoproduction data~\cite{GDH:2001zzk,GDH:2003xhc,Dutz:2004zz} in the region of the $\Delta(1232)$, but there is a $4\sigma$ disagreement in the second resonance region based on the total uncertainty and assuming uncorrelated systematic uncertainty. 
Our deuteron results show a higher $\Delta(1232)$ than the photoproduction data~\cite{Ahrens:2006yx, Ahrens:2009zz}, though the difference is only at the $2\sigma$ level. 
\begin{table}[!ht]
\caption{Integrals of $\sigma_{TT}$ for the proton, deuteron and $^3$He shown in Figs.~\ref{fig:sigtt_pd3He} over the energy ranges for the $\Delta(1232)$ ($0.22\,\textrm{GeV}\leqslant \nu\leqslant 0.5\,\textrm{GeV}$),
second ($0.5\,\textrm{GeV}\leqslant\nu\leqslant 0.88\,\textrm{GeV}$) and third resonance regions ($0.88\,\textrm{GeV}\leqslant\nu\leqslant 1.18\,\textrm{GeV}$), in unit $\mu$b$\cdot$GeV.
The data are from either photoproduction $(\gamma)$ or electroproduction data extrapolated to $Q^2=0$ $(e^-)$. 
The uncertainties are the quadratic sum of statistical and systematic errors.  Values shown in {\it italic} fonts indicate disagreements exceeding $3\sigma$ level between datasets.  
}
\label{tab:integ}
\begin{center}
\begin{tabular}{ l l c c c  }
\hline
Target & Data type & $\Delta(1232)$ & 2nd & 3rd \\ 
\hline
$p$ & $(\gamma)$
 & {62.0~$\pm$~1.7}  & \textit{31.9~$\pm$~1.3}  & {17.5~$\pm$~1.3} \\
 & $(e^-)$
 & {63.8~$\pm$~2.8}  & \textit{21.8~$\pm$~2.2}  & {20.2~$\pm$~5.4} \\
\hline
$d$ & $(\gamma)$
 & {106.4~$\pm$~2.6}  & {56.3~$\pm$~3.2}  & {33.9~$\pm$~4.6} \\
 & $(e^-)$
 & {130.2~$\pm$~12.0}  & {49.0~$\pm$~5.3}  & {29.9~$\pm$~3.8} \\
\hline
$^3$He & $(\gamma)$ & {40.8~$\pm$~5.8} & $-$ & $-$\\
   & $(e^-)$
 & {62.7~$\pm$~2.8}  & {13.7~$\pm$~1.6}  & {10.7~$\pm$~1.9} \\
\hline
\end{tabular}
\end{center}
\end{table}
\section{Extraction of neutron information using WBA \label{sec:WBA}}

The WBA method constructs unpolarized and polarized structure functions of the light nuclei ($A=2,3$) by convoluting those of the proton and the neutron with smearing functions calculated from the nuclear wavefunctions~\cite{Kahn:2008nq,Ethier:2013hna,Ethier:2014bua,Tropiano:2018quk}. 
The WBA method was used to extract neutron $g_1^n$ from the $g_1^p$ and $g_1^d$ results of EG4 experiment~\cite{Deur:2021klh}. {The basic procedure is as follows: we first subtract the proton contribution from the deuteron to obtain the $g_1$ of the bound neutron. We then start with an input $g_1^n$ model~\cite{CLAS:2015otq}, apply WBA smearing and compare the smeared model with the bound neutron data. The input model is adjusted until an agreement is reached, effectively ``unsmearing" the measured deuteron data.} 

\begin{figure}[!htb]
\includegraphics[width=0.49\textwidth]{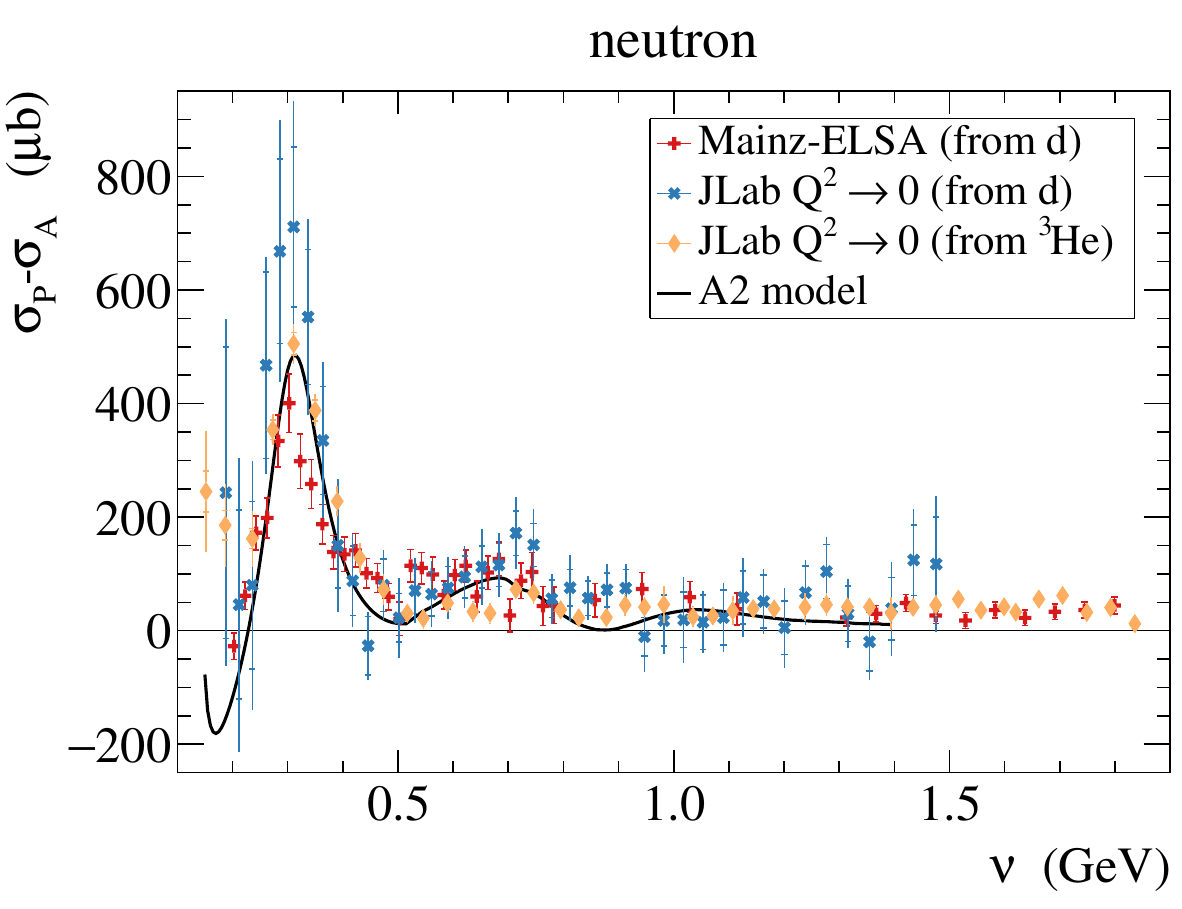}
\caption{$\sigma_{TT}^n$ for the neutron extracted from either photoproduction data of the deuteron (red $+$), or electroproduction data of the deuteron (blue $\times$) and $^{3}$He  (orange diamonds) extrapolated to $Q^2=0$. Also shown is a model prediction, see text for details.  
}
\label{fig:sigtt_n}
\end{figure}

The WBA unsmearing procedure was applied here to extract the neutron $\sigma_{TT}$ from  deuteron data, obtained from both  photoproduction~\cite{Ahrens:2006yx, Ahrens:2009zz} and electroproduction channels. 
For electroproduction data of EG4~\cite{Deur:2021klh}, we first extrapolated $A_1F_1$ of the proton and the deuteron to $Q^2=0$, then extracted $A_1F_1$ of the neutron using WBA unsmearing and formed $\sigma_{TT}^n$ using Eq.~(\ref{eq:sigtt-A1F1}). Furthermore, the WBA unsmearing procedure was applied to $^3$He for the first time and we extracted $\sigma_{TT}^n$ from the extrapolated electroproduction data of $^3$He (Fig.~\ref{fig:sigtt_pd3He} right panel). 
In both the deuteron and $^3$He unsmearing, the corresponding proton data from photoproduction~\cite{
GDH:2001zzk,GDH:2003xhc,Dutz:2004zz} or electroproduction~\cite{CLAS:2021apd} were used as inputs to account for the proton contribution. 
For $^3$He, the proton data were supplemented with the SAID model~\cite{Workman:2012jf,Briscoe:2023gmb} for the two lowest $\nu$ bins.
The uncertainty of the extraction was studied by varying the initial input model to the WBA unsmearing procedure, and by varying the deuteron or $^3$He wavefunctions used to calculate the smearing functions. Both uncertainties were found to be negligible compared with the uncertainty of the data themselves. 
The method of fitting, extrapolating to $Q^2=0$, and uncertainty estimation is the same as that in Sec.~\ref{sec:pdhe}. 
The $\sigma_{TT}^n$ from all three channels are shown in Fig.~\ref{fig:sigtt_n} together with a model constructed~\cite{Pedroni:2026eqj} as a sum of final states $N\pi$~\cite{Strakovsky:2022tvu}, $N\pi\pi$~\cite{Fix:2005if}, and $N\eta$~\cite{Tiator:2018heh}.

\begin{table}[!h]
\caption{Same as Table~\ref{tab:integ} but for $\sigma_{TT}^n$ (Fig.~\ref{fig:sigtt_n}), divided additionally into whether the information was extracted from the deuteron or $^3$He. 
Values shown in {\bf bold} ({\it italic}) fonts indicate disagreements exceeding $4\sigma$ ($3\sigma$) level between datasets. 
}
\label{tab:integ2}
\begin{center}
\begin{tabular}{ l l c c c  }
\hline
 & Data type & $\Delta(1232)$ & 2nd & 3rd \\ 
\hline
$n$ & $(\gamma~d)$
 & \textbf{51.3~$\pm$~2.6}  & \textit{28.6~$\pm$~3.2}  & {16.5~$\pm$~4.6} \\
 & $(e^-~d)$
 & {80.6~$\pm$~13.2}  &  {34.5~$\pm$~6.0}  & {10.0~$\pm$~6.2} \\
 & $(e^-~^3$He)
 & \textbf{71.5~$\pm$~3.3}  & \textit{15.1~$\pm$~1.9}  & {10.8~$\pm$~2.3} \\
\hline
\end{tabular}
\end{center}
\end{table}

The three data sets were compared using integrals over the three resonance regions, see Table~\ref{tab:integ2}. Disagreement exceeding $3\sigma$ level are seen in both the $\Delta(1232)$ and the second resonance regions. Specifically, there is a $4.8\sigma$ difference between the extraction from photoproduction on deuteron and electroproduction on $^3$He in the $\Delta(1232)$ region.

Remarkably, neutron results extracted from the deuteron and $^3$He electroproduction data are consistent, both resulting in a $\sigma_{\rm TT}^n$ similar in magnitude to $\sigma_{\rm TT}^p$ in the $\Delta(1232)$ region, as expected from isospin symmetry~\cite{Burkert:2000qm}. 
This may open the path of applying the WBA framework to extract neutron information from a wealth of $^3$He data at real and quasi-real photon (zero and very low $Q^2$) kinematics~\cite{Amarian:2002ar, JeffersonLabE97-110:2019fsc, E01-012:2013fky, JeffersonLabHallA:2016neg},  which has not yet been attempted.

\section{Isovector Results}
\begin{figure} [!ht]
\centering
\includegraphics[width=0.49\textwidth]{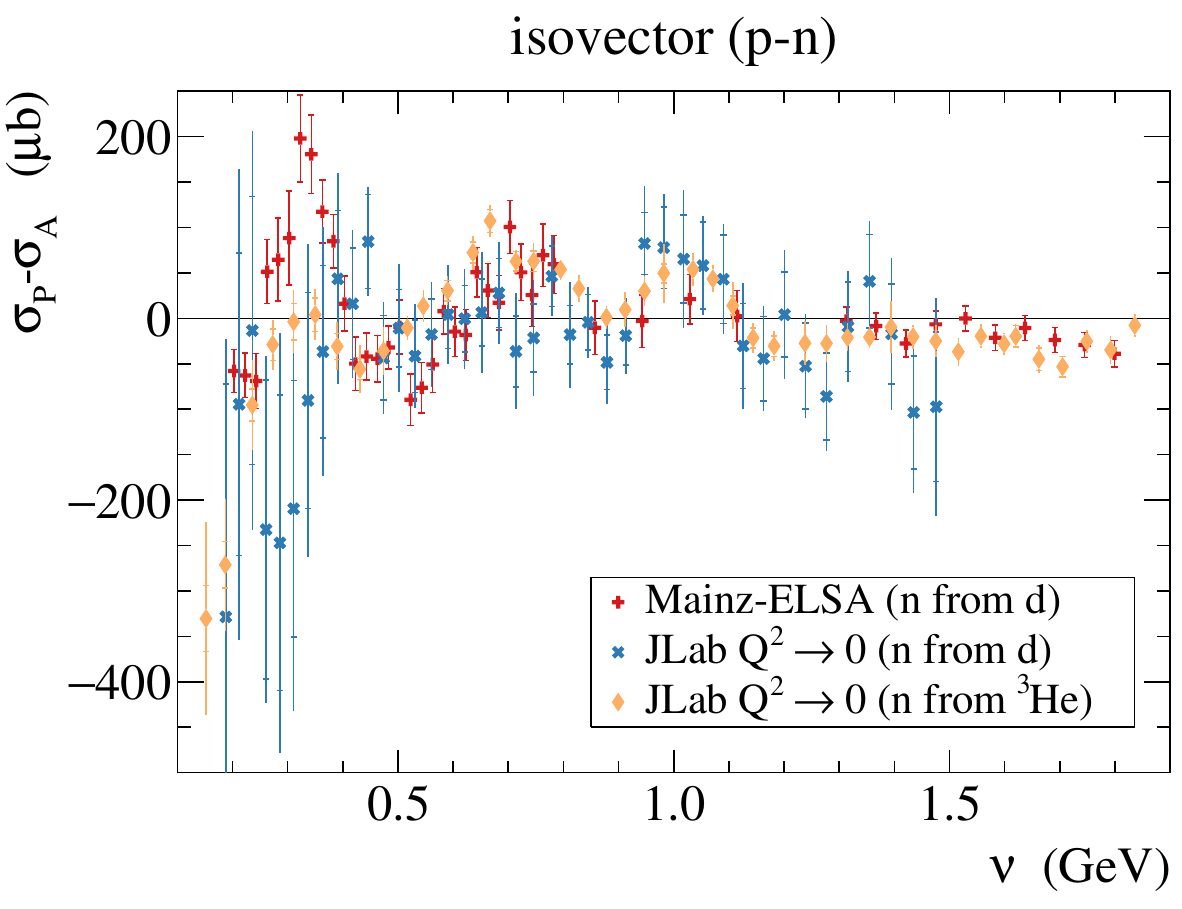}
\caption{Isovector combination $\sigma_{TT}^{p-n}(\nu)$ from three different combinations of photo- vs. electro-production, and deuteron vs. $^3$He data. }
    \label{fig:isovector}
\end{figure}

A convenient way to compare the proton and neutron results is through the isovector difference, $\sigma^{p-n}_{\rm TT}$. In  Fig.~\ref{fig:isovector}, we show results for $\sigma^{p-n}_{\rm TT}(\nu)$ where the neutron data were obtained from photo- and electroproductions from deuteron or electroproduction from $^3$He.  For the proton input, we used the data from Mainz-ELSA as they are the most precise in the full $\nu$ region and are consistent with electroproduction data at least in the $\Delta(1232)$ region.
As with the other data we integrate the resonance regions, see Table.~\ref{tab:integ3}. 
\begin{table}[!h]
\caption{Same as Table~\ref{tab:integ} but for $\sigma_{TT}^{p-n}$ results of Fig.~\ref{fig:isovector}, divided into whether the neutron information was extracted from photo- or electroproduction, and from the deuteron or $^3$He. 
Values shown in bold fonts indicate disagreements exceeding $3\sigma$ level among datasets. 
}
\label{tab:integ3}
\begin{center}
\begin{tabular}{ l l c c c  }
\hline
 & Data type & $\Delta(1232)$ & 2nd & 3rd \\ 
 \hline
$p-n$ & $(\gamma~d)$
 & \textbf{10.4~$\pm$~3.1}  & {4.1~$\pm$~3.5}  & {1.6~$\pm$~5.0} \\
 & $(e^-~d)$
 & {-18.8~$\pm$~13.3}  & {-2.7~$\pm$~6.2}  & {7.5~$\pm$~6.3} \\
 & $(e^-~^3$He)
 & \textbf{-9.5~$\pm$~4.0}  & \textbf{17.1~$\pm$~2.6}  & {6.7~$\pm$~2.6} \\
 \hline
\end{tabular}
\end{center}
\end{table}

From the known isospin-symmetric $p\leftrightarrow n$ structure of the $N\to\Delta$ transition, we expect $\sigma_{TT}^{p-n} = 0$ at the $\Delta(1232)$~\cite{Burkert:2000qm}, although this can be obscured by non-resonant background. Other non-zero signals may also come from final state interactions and meson exchange currents, as well as elastic, quasi-elastic, and higher mass resonances reaching under the $\Delta$ due to experimental resolution or imperfect radiative corrections. 
However, these non-zero contributions become relatively suppressed at the $\Delta(1232)$ peak.  
Indeed, near the center of the peak, $\nu=0.34$ GeV, the JLab-based data point are compatible with zero. 
The Mainz-ELSA data, however, show a non-zero signal. 
More generally, Eq.~(\ref{eq:gdh}) suggests that in average, $\sigma_{TT}^{p-n}$ is small because 
while $\kappa$ is different for proton ($+1.793$) and neutron ($-1.913$), their $\kappa^2/M^2$ differ by only about 13\%. Therefore, the isovector GDH sum is only about 13\% of the magnitude of the individual nucleon GDH sum. 
Notably, the isovector combination $I_{\rm GDH}^{p-n}$ is significant because of its connection to the Bjorken sum rule~\cite{Bjorken:1968dy,Deur:2018roz}:  $I_{\rm GDH}^{p-n} = \frac{16\alpha \pi^2}{Q^2} \Gamma_1^{p-n}$ with $\Gamma_1^{p-n}$ the Bjorken integral. In particular, the latter provides the QCD effective coupling~\cite{Deur:2023dzc} $\alpha_{g_1}=\pi\big(1-\frac{6}{g_A}\Gamma_1^{p-n} \big)$.
The smallness of $(\kappa_p^2/M_p^2-\kappa_n^2/M_n^2)$ thus implies a small slope of $\alpha_{g_1}$ at $Q^2=0$, providing nontrivial dynamical information on the strong force in the long distance regime, namely the freezing of the QCD coupling~\cite{Deur:2018roz}. 
In turn, since the freezing of $\alpha_{g_1}$ is governed by confinement dynamics~\cite{Brodsky:2008be}, the observed smallness of $I_{\rm GDH}^{p-n}$ may thus be viewed as a manifestation of confinement, providing a complementary perspective on $\kappa_p^2 \simeq \kappa_n^2$ to that stemming from isospin symmetry.

\section{Conclusion}

In this article, we extrapolated very low $Q^2$ electroproduction $\sigma_{TT}$ results for the proton, deuteron, and $^3$He to the $Q^2=0$ real photon point.  This provides independent data with different types of systematic uncertainties, and for the $^3$He case also significantly higher precision and energy coverage, compared to photoproduction data. 
The extrapolated electroproduction results show similar $\Delta(1232)$ strength between the proton and neutron extracted from either deuteron or $^3$He, as expected from isospin symmetry. This may help to understand the surprising result from the photoproduction measurements where the $\Delta(1232)$ from deuteron is significantly smaller than twice that from proton. 
Furthermore, we applied the weak binding approximation to $^3$He data at very low $Q^2$ for the first time, with encouraging results regarding the possibility of extracting neutron spin structure functions from $^3$He data outside the deep inelastic scattering region. 
\section*{Acknowledgments}
We thank J.-P.~Chen, S.E.~Kuhn and W.~Melnitchouk for useful discussions and P. Pedroni for information about the preliminary A2 data. 
This material is based upon work supported by the U.S. Department of Energy, Office of Science, Office of Nuclear Physics under Contract No. 89243126CSC000213.
The work of D. W. Upton was supported by the U.S. Department of Energy, Office of Science, Office of Nuclear Physics under contract number DE-FG02-96ER40960, and the Jefferson Science Associates Graduate Fellowship. 
The work of Y.~Li and X.~Zheng was supported by the U.S. Department of Energy, Office of Science, Office of Nuclear Physics under contract number DE–SC0014434. The work of B.~Callahan, O.~Larson and A.~Rask was performed to fulfill the undergraduate research requirement of the Department of Physics at the University of Virginia. 

\appendix

\section{Data Tables}
\label{sec:data}

Tables contain the extrapolated data as described in the text.

\begin{table}[htb]
\label{tab:p}
\begin{center}
\caption{Cross sections extrapolated to $Q^2=0$ for $^1$H.}
\begin{tabular}{ c | c | S[table-format=4.2] | c | c}
 $W$ & $\nu$ & $\sigma_{TT}$ & stat & syst \\ 
 GeV & GeV & $\mu$b  & $\mu$b & $\mu$b \\ 
 \hline
1.11 & 0.188 & -93.8 & 37.0 & 41.9 \\ 
1.13 & 0.212 & -39.7 & 32.6 & 37.5 \\ 
1.15 & 0.236 & 101.1 & 33.5 & 18.5 \\ 
1.17 & 0.261 & 262.5 & 34.8 & 28.2 \\ 
1.19 & 0.286 & 449.9 & 34.2 & 31.0 \\ 
1.21 & 0.311 & 524.4 & 30.5 & 30.6 \\ 
1.23 & 0.337 & 476.9 & 27.9 & 22.8 \\ 
1.25 & 0.364 & 317.7 & 23.5 & 15.7 \\ 
1.27 & 0.391 & 189.3 & 19.1 & 12.9 \\ 
1.29 & 0.418 & 77.0 & 16.9 & 13.4 \\ 
1.31 & 0.446 & 48.6 & 15.3 & 14.1 \\ 
1.33 & 0.474 & 8.2 & 15.1 & 10.5 \\ 
1.35 & 0.502 & 22.6 & 14.8 & 11.3 \\ 
1.37 & 0.531 & 31.8 & 14.3 & 11.4 \\ 
1.39 & 0.561 & 28.2 & 12.8 & 12.9 \\ 
1.41 & 0.591 & 50.7 & 12.6 & 16.4 \\ 
1.43 & 0.621 & 79.9 & 12.6 & 11.8 \\ 
1.45 & 0.652 & 80.1 & 12.9 & 14.5 \\ 
1.47 & 0.683 & 88.6 & 15.2 & 21.6 \\ 
1.49 & 0.714 & 74.3 & 15.8 & 20.4 \\ 
1.51 & 0.746 & 92.2 & 15.3 & 11.3 \\ 
1.53 & 0.779 & 48.1 & 14.6 & 11.6 \\ 
1.55 & 0.812 & 67.9 & 14.6 & 13.6 \\ 
1.57 & 0.845 & 30.5 & 15.1 & 10.9 \\ 
1.59 & 0.879 & 13.5 & 15.2 & 16.2 \\ 
1.61 & 0.913 & 34.3 & 15.8 & 17.3 \\ 
1.63 & 0.947 & 43.9 & 23.1 & 30.2 \\ 
1.65 & 0.982 & 44.4 & 38.0 & 49.0 \\ 
1.67 & 1.018 & 163.4 & 38.3 & 61.1 \\ 
1.69 & 1.053 & 86.8 & 36.7 & 27.4 \\ 
1.71 & 1.090 & 88.9 & 36.2 & 52.9 \\ 
1.73 & 1.126 & 69.9 & 34.6 & 41.8 \\ 
1.75 & 1.163 & 27.4 & 33.3 & 38.6 \\ 
1.77 & 1.201 & 51.4 & 33.6 & 45.3 \\ 
1.79 & 1.239 & 16.0 & 33.7 & 46.6 \\ 
1.81 & 1.277 & -4.1 & 34.4 & 45.3 \\ 
1.83 & 1.316 & 36.1 & 35.0 & 27.2 \\ 
1.85 & 1.355 & -5.9 & 35.8 & 45.3 \\ 
1.87 & 1.395 & 89.5 & 36.1 & 58.0 \\ 
1.89 & 1.435 & -34.5 & 36.7 & 55.0 \\ 
1.91 & 1.476 & 44.9 & 38.2 & 38.9 \\ 
1.93 & 1.517 & 58.8 & 45.6 & 67.2 \\ 
1.95 & 1.558 & 18.0 & 51.7 & 8.5 \\ 
1.97 & 1.600 & 37.5 & 53.8 & 116.0 \\ 
1.99 & 1.642 & 1.8 & 50.4 & 155.6 \\ 
2.01 & 1.685 & -11.1 & 55.3 & 91.9 \\ 
2.03 & 1.728 & 45.5 & 61.4 & 56.2 \\ 
\end{tabular}
\end{center}
\end{table}

\begin{table}[htb]
\label{tab:d}
\begin{center}
\caption{Cross sections extrapolated to $Q^2=0$ for $^2$H.}
\begin{tabular}{ c | c | S[table-format=4.2] | c | c}
 $W$ & $\nu$ & $\sigma_{TT}$ & stat & syst \\ 
 GeV & GeV & $\mu$b  & $\mu$b & $\mu$b \\ 
 \hline
1.11 & 0.188 & 236.4 & 202.7 & 212.2 \\ 
1.13 & 0.212 & 49.9 & 158.1 & 189.7 \\ 
1.15 & 0.236 & 176.6 & 138.1 & 154.1 \\ 
1.17 & 0.261 & 670.5 & 149.4 & 95.2 \\ 
1.19 & 0.286 & 969.5 & 141.5 & 153.0 \\ 
1.21 & 0.311 & 1067.8 & 120.4 & 147.2 \\ 
1.23 & 0.337 & 894.4 & 104.3 & 108.6 \\ 
1.25 & 0.364 & 598.4 & 87.0 & 93.2 \\ 
1.27 & 0.391 & 322.9 & 70.3 & 80.8 \\ 
1.29 & 0.418 & 180.0 & 57.4 & 49.8 \\ 
1.31 & 0.446 & 23.3 & 48.1 & 29.2 \\ 
1.33 & 0.474 & 100.4 & 42.7 & 37.6 \\ 
1.35 & 0.502 & 38.4 & 38.7 & 51.6 \\ 
1.37 & 0.531 & 85.1 & 36.3 & 35.0 \\ 
1.39 & 0.561 & 85.1 & 35.2 & 39.1 \\ 
1.41 & 0.591 & 107.4 & 34.1 & 32.8 \\ 
1.43 & 0.621 & 140.8 & 33.3 & 40.4 \\ 
1.45 & 0.652 & 164.6 & 33.0 & 51.9 \\ 
1.47 & 0.683 & 169.9 & 32.5 & 35.2 \\ 
1.49 & 0.714 & 211.7 & 31.8 & 39.5 \\ 
1.51 & 0.746 & 190.0 & 30.9 & 44.5 \\ 
1.53 & 0.779 & 106.2 & 28.7 & 24.3 \\ 
1.55 & 0.812 & 124.0 & 28.5 & 45.2 \\ 
1.57 & 0.845 & 93.3 & 26.9 & 20.8 \\ 
1.59 & 0.879 & 102.0 & 25.6 & 27.4 \\ 
1.61 & 0.913 & 104.2 & 27.0 & 23.0 \\ 
1.63 & 0.947 & 31.0 & 26.3 & 40.2 \\ 
1.65 & 0.982 & 69.9 & 24.6 & 22.9 \\ 
1.67 & 1.018 & 143.9 & 24.9 & 34.4 \\ 
1.69 & 1.053 & 129.2 & 23.2 & 24.3 \\ 
1.71 & 1.090 & 90.2 & 23.1 & 20.3 \\ 
1.73 & 1.126 & 137.7 & 23.4 & 27.1 \\ 
1.75 & 1.163 & 86.3 & 23.1 & 25.6 \\ 
1.77 & 1.201 & 53.6 & 23.2 & 29.9 \\ 
1.79 & 1.239 & 93.7 & 23.0 & 19.9 \\ 
1.81 & 1.277 & 89.3 & 23.0 & 28.9 \\ 
1.83 & 1.316 & 76.3 & 23.3 & 23.7 \\ 
1.85 & 1.355 & 57.0 & 23.5 & 26.6 \\ 
1.87 & 1.395 & 52.4 & 24.9 & 31.2 \\ 
1.89 & 1.435 & 109.3 & 26.5 & 24.8 \\ 
1.91 & 1.476 & 86.9 & 30.7 & 32.8 
\end{tabular}
\end{center}
\end{table}

\begin{table}[htb]
\label{tab:n}
\begin{center}
\caption{Cross sections extrapolated to $Q^2=0$ for n.}
\begin{tabular}{ c | c | S[table-format=4.2] | c | c}
 $W$ & $\nu$ & $\sigma_{TT}$ & stat & syst \\ 
 GeV & GeV & $\mu$b  & $\mu$b & $\mu$b \\ 
 \hline
1.11 & 0.188 & 243.1 & 256.6 & 165.6 \\ 
1.13 & 0.212 & 45.9 & 166.3 & 198.5 \\ 
1.15 & 0.236 & 80.1 & 147.6 & 162.0 \\ 
1.17 & 0.261 & 467.4 & 164.6 & 96.9 \\ 
1.19 & 0.286 & 668.0 & 162.6 & 163.8 \\ 
1.21 & 0.311 & 711.1 & 141.2 & 172.1 \\ 
1.23 & 0.337 & 552.5 & 118.8 & 124.3 \\ 
1.25 & 0.364 & 335.0 & 94.8 & 99.4 \\ 
1.27 & 0.391 & 150.0 & 75.0 & 89.0 \\ 
1.29 & 0.418 & 87.7 & 61.2 & 54.2 \\ 
1.31 & 0.446 & -26.6 & 51.7 & 29.9 \\ 
1.33 & 0.474 & 80.1 & 46.5 & 40.7 \\ 
1.35 & 0.502 & 22.8 & 42.5 & 55.8 \\ 
1.37 & 0.531 & 70.6 & 40.0 & 40.4 \\ 
1.39 & 0.561 & 64.3 & 38.8 & 43.2 \\ 
1.41 & 0.591 & 75.3 & 37.5 & 40.0 \\ 
1.43 & 0.621 & 94.6 & 36.8 & 40.6 \\ 
1.45 & 0.652 & 112.5 & 36.9 & 55.3 \\ 
1.47 & 0.683 & 115.7 & 38.0 & 41.7 \\ 
1.49 & 0.714 & 171.9 & 39.2 & 50.3 \\ 
1.51 & 0.746 & 151.1 & 37.7 & 51.8 \\ 
1.53 & 0.779 & 56.1 & 33.2 & 28.8 \\ 
1.55 & 0.812 & 75.7 & 32.1 & 48.7 \\ 
1.57 & 0.845 & 57.4 & 30.6 & 24.0 \\ 
1.59 & 0.879 & 71.8 & 30.0 & 34.6 \\ 
1.61 & 0.913 & 75.0 & 32.5 & 26.5 \\ 
1.63 & 0.947 & -10.7 & 34.0 & 53.0 \\ 
1.65 & 0.982 & 17.9 & 44.9 & 38.4 \\ 
1.67 & 1.018 & 19.0 & 48.7 & 58.5 \\ 
1.69 & 1.053 & 14.9 & 47.9 & 26.0 \\ 
1.71 & 1.090 & 23.0 & 48.6 & 36.6 \\ 
1.73 & 1.126 & 58.4 & 46.7 & 51.5 \\ 
1.75 & 1.163 & 51.3 & 46.4 & 34.4 \\ 
1.77 & 1.201 & 5.2 & 46.8 & 53.1 \\ 
1.79 & 1.239 & 67.0 & 47.3 & 32.2 \\ 
1.81 & 1.277 & 104.1 & 47.7 & 36.9 \\ 
1.83 & 1.316 & 29.9 & 49.1 & 36.4 \\ 
1.85 & 1.355 & -19.7 & 51.7 & 42.5 \\ 
1.87 & 1.395 & 38.5 & 55.2 & 62.4 \\ 
1.89 & 1.435 & 124.4 & 62.1 & 64.0 \\ 
1.91 & 1.476 & 117.4 & 82.4 & 87.5 
\end{tabular}
\end{center}
\end{table}

\begin{table}[htb]
\label{tab:3He}
\begin{center}
\caption{Cross sections extrapolated to $Q^2=0$ for $^3$He.}
\begin{tabular}{ c | c | S[table-format=3.2] | c | c}
 $W$ & $\nu$ & $\sigma_{TT}$ & stat & syst \\ 
 GeV & GeV & $\mu$b  & $\mu$b & $\mu$b \\ 
 \hline
1.08 & 0.152 & 245.0 & 36.4 & 99.6 \\ 
1.11 & 0.187 & 185.8 & 25.7 & 68.2 \\ 
1.15 & 0.236 & 162.1 & 17.5 & 46.2 \\ 
1.18 & 0.273 & 353.9 & 17.4 & 21.3 \\ 
1.21 & 0.311 & 505.3 & 19.9 & 28.6 \\ 
1.24 & 0.350 & 388.0 & 18.5 & 21.0 \\ 
1.27 & 0.390 & 227.9 & 14.2 & 22.2 \\ 
1.30 & 0.431 & 128.0 & 11.7 & 23.6 \\ 
1.33 & 0.474 & 72.3 & 10.3 & 24.7 \\ 
1.36 & 0.517 & 30.6 & 8.4 & 10.1 \\ 
1.38 & 0.546 & 21.6 & 10.1 & 14.1 \\ 
1.41 & 0.590 & 48.7 & 11.4 & 11.1 \\ 
1.44 & 0.636 & 33.6 & 11.6 & 14.8 \\ 
1.46 & 0.667 & 30.8 & 12.6 & 12.8 \\ 
1.49 & 0.714 & 72.7 & 10.7 & 6.5 \\ 
1.51 & 0.746 & 66.7 & 11.3 & 16.7 \\ 
1.54 & 0.795 & 36.8 & 9.7 & 4.6 \\ 
1.56 & 0.828 & 22.7 & 10.3 & 10.5 \\ 
1.59 & 0.878 & 23.3 & 8.4 & 9.1 \\ 
1.61 & 0.912 & 45.3 & 10.0 & 16.7 \\ 
1.63 & 0.947 & 41.7 & 10.1 & 13.4 \\ 
1.65 & 0.982 & 46.1 & 10.5 & 30.8 \\ 
1.68 & 1.035 & 24.5 & 8.9 & 16.1 \\ 
1.70 & 1.071 & 26.0 & 9.3 & 11.5 \\ 
1.72 & 1.107 & 35.1 & 10.5 & 23.3 \\ 
1.74 & 1.144 & 38.7 & 11.1 & 12.4 \\ 
1.76 & 1.182 & 38.4 & 11.4 & 11.9 \\ 
1.79 & 1.238 & 41.5 & 9.1 & 22.8 \\ 
1.81 & 1.277 & 45.9 & 9.1 & 18.4 \\ 
1.83 & 1.315 & 41.7 & 9.2 & 12.3 \\ 
1.85 & 1.355 & 41.8 & 10.0 & 6.8 \\ 
1.87 & 1.394 & 31.1 & 10.2 & 26.4 \\ 
1.89 & 1.434 & 41.2 & 10.3 & 8.2 \\ 
1.91 & 1.475 & 45.2 & 10.0 & 14.6 \\ 
1.93 & 1.516 & 55.5 & 9.8 & 11.9 \\ 
1.95 & 1.557 & 35.9 & 9.8 & 8.3 \\ 
1.97 & 1.599 & 42.0 & 9.8 & 6.9 \\ 
1.98 & 1.620 & 32.6 & 11.9 & 5.5 \\ 
2.00 & 1.662 & 55.7 & 12.3 & 9.0 \\ 
2.02 & 1.705 & 62.0 & 11.3 & 2.0 \\ 
2.04 & 1.749 & 32.1 & 10.3 & 7.0 \\ 
2.06 & 1.792 & 40.7 & 9.9 & 12.2 \\ 
2.08 & 1.836 & 12.4 & 10.4 & 7.5 
\end{tabular}
\end{center}
\end{table}

\begin{table}[htb]
\label{tab:n_3He}
\begin{center}
\caption{Cross sections for neutron extracted from $^3$He}
\begin{tabular}{ c | c | S[table-format=3.2] | c | c}
 $W$ & $\nu$ & $\sigma_{TT}$ & stat & syst \\ 
 GeV & GeV & $\mu$b  & $\mu$b & $\mu$b \\ 
 \hline
1.08 & 0.152 & 245.0 & 18.2 & 49.8 \\ 
1.11 & 0.187 & 185.8 & 12.8 & 34.1 \\ 
1.15 & 0.236 & 162.1 & 8.7 & 23.1 \\ 
1.18 & 0.273 & 353.9 & 8.7 & 10.6 \\ 
1.21 & 0.311 & 505.3 & 9.9 & 14.3 \\ 
1.24 & 0.350 & 388.0 & 9.3 & 10.5 \\ 
1.27 & 0.390 & 227.9 & 7.1 & 11.1 \\ 
1.30 & 0.431 & 128.0 & 5.8 & 11.8 \\ 
1.33 & 0.474 & 72.3 & 5.1 & 12.3 \\ 
1.36 & 0.517 & 30.6 & 4.2 & 5.1 \\ 
1.38 & 0.546 & 21.6 & 5.0 & 7.0 \\ 
1.41 & 0.590 & 48.7 & 5.7 & 5.6 \\ 
1.44 & 0.636 & 33.6 & 5.8 & 7.4 \\ 
1.46 & 0.667 & 30.8 & 6.3 & 6.4 \\ 
1.49 & 0.714 & 72.7 & 5.4 & 3.3 \\ 
1.51 & 0.746 & 66.7 & 5.7 & 8.4 \\ 
1.54 & 0.795 & 36.8 & 4.8 & 2.3 \\ 
1.56 & 0.828 & 22.7 & 5.1 & 5.3 \\ 
1.59 & 0.878 & 23.3 & 4.2 & 4.5 \\ 
1.61 & 0.912 & 45.3 & 5.0 & 8.3 \\ 
1.63 & 0.947 & 41.7 & 5.1 & 6.7 \\ 
1.65 & 0.982 & 46.1 & 5.2 & 15.4 \\ 
1.68 & 1.035 & 24.5 & 4.4 & 8.0 \\ 
1.70 & 1.071 & 26.0 & 4.7 & 5.7 \\ 
1.72 & 1.107 & 35.1 & 5.2 & 11.6 \\ 
1.74 & 1.144 & 38.7 & 5.6 & 6.2 \\ 
1.76 & 1.182 & 38.4 & 5.7 & 5.9 \\ 
1.79 & 1.238 & 41.5 & 4.5 & 11.4 \\ 
1.81 & 1.277 & 45.9 & 4.5 & 9.2 \\ 
1.83 & 1.315 & 41.7 & 4.6 & 6.1 \\ 
1.85 & 1.355 & 41.8 & 5.0 & 3.4 \\ 
1.87 & 1.394 & 31.1 & 5.1 & 13.2 \\ 
1.89 & 1.434 & 41.2 & 5.2 & 4.1 \\ 
1.91 & 1.475 & 45.2 & 5.0 & 7.3 \\ 
1.93 & 1.516 & 55.5 & 4.9 & 6.0 \\ 
1.95 & 1.557 & 35.9 & 4.9 & 4.2 \\ 
1.97 & 1.599 & 42.0 & 4.9 & 3.5 \\ 
1.98 & 1.620 & 32.6 & 6.0 & 2.7 \\ 
2.00 & 1.662 & 55.7 & 6.1 & 4.5 \\ 
2.02 & 1.705 & 62.0 & 5.7 & 1.0 \\ 
2.04 & 1.749 & 32.1 & 5.1 & 3.5 \\ 
2.06 & 1.792 & 40.7 & 4.9 & 6.1 \\ 
2.08 & 1.836 & 12.4 & 5.2 & 3.8 
\end{tabular}
\end{center}
\end{table}

\bibliographystyle{elsarticle-num}
\bibliography{references.bib}

\begin{thebibliography}{10}
\expandafter\ifx\csname url\endcsname\relax
  \def\url#1{\texttt{#1}}\fi
\expandafter\ifx\csname urlprefix\endcsname\relax\def\urlprefix{URL }\fi
\expandafter\ifx\csname href\endcsname\relax
  \def\href#1#2{#2} \def\path#1{#1}\fi

\bibitem{EuropeanMuon:1987isl}
J.~Ashman, et~al., {A Measurement of the Spin Asymmetry and Determination of the Structure Function g(1) in Deep Inelastic Muon-Proton Scattering}, Phys. Lett. B 206 (1988) 364.
\newblock \href {https://doi.org/10.1016/0370-2693(88)91523-7} {\path{doi:10.1016/0370-2693(88)91523-7}}.

\bibitem{Anselmino:1994gn}
M.~Anselmino, A.~Efremov, E.~Leader, The theory and phenomenology of polarized deep inelastic scattering, Phys. Rept. 261 (1995) 1--124, [Erratum: Phys.Rept. 281, 399--400 (1997)].
\newblock \href {http://arxiv.org/abs/hep-ph/9501369} {\path{arXiv:hep-ph/9501369}}, \href {https://doi.org/10.1016/0370-1573(95)00011-5} {\path{doi:10.1016/0370-1573(95)00011-5}}.

\bibitem{Deur:2018roz}
A.~Deur, S.~J. Brodsky, G.~F. De~T\'eramond, {The Spin Structure of the Nucleon}, Rept. Prog. Phys. 82 (2019) 076201.
\newblock \href {http://arxiv.org/abs/1807.05250} {\path{arXiv:1807.05250}}, \href {https://doi.org/10.1088/1361-6633/ab0b8f} {\path{doi:10.1088/1361-6633/ab0b8f}}.

\bibitem{Ji:2020ena}
X.~Ji, F.~Yuan, Y.~Zhao, {What we know and what we don\textquoteright{}t know about the proton spin after 30 years}, Nature Rev. Phys. 3~(1) (2021) 27--38.
\newblock \href {http://arxiv.org/abs/2009.01291} {\path{arXiv:2009.01291}}, \href {https://doi.org/10.1038/s42254-020-00248-4} {\path{doi:10.1038/s42254-020-00248-4}}.

\bibitem{Gerasimov:1965et}
S.~B. Gerasimov, A sum rule for magnetic moments and the damping of the nucleon magnetic moment in nuclei, Yad. Fiz. 2 (1965) 598--602.

\bibitem{Drell:1966jv}
S.~D. Drell, A.~C. Hearn, Exact sum rule for nucleon magnetic moments, Phys. Rev. Lett. 16 (1966) 908--911.
\newblock \href {https://doi.org/10.1103/PhysRevLett.16.908} {\path{doi:10.1103/PhysRevLett.16.908}}.

\bibitem{Altarelli:1972nc}
G.~Altarelli, N.~Cabibbo, L.~Maiani, {The Drell-Hearn sum rule and the lepton magnetic moment in the Weinberg model of weak and electromagnetic interactions}, Phys. Lett. B 40 (1972) 415--419.
\newblock \href {https://doi.org/10.1016/0370-2693(72)90833-7} {\path{doi:10.1016/0370-2693(72)90833-7}}.

\bibitem{Dalton:2020wdv}
M.~M. Dalton, A.~Deur, C.~D. Keith, S.~\v{S}irca, J.~Stevens, {Measurement of the high-energy contribution to the Gerasimov-Drell-Hearn sum rule} (8 2020).
\newblock \href {http://arxiv.org/abs/2008.11059} {\path{arXiv:2008.11059}}.

\bibitem{GDH:2001zzk}
J.~Ahrens, et~al., First measurement of the {G}erasimov-{D}rell-{H}earn integral for hydrogen from 200 to 800 {MeV}, Phys. Rev. Lett. 87 (2001) 022003.
\newblock \href {http://arxiv.org/abs/hep-ex/0105089} {\path{arXiv:hep-ex/0105089}}, \href {https://doi.org/10.1103/PhysRevLett.87.022003} {\path{doi:10.1103/PhysRevLett.87.022003}}.

\bibitem{GDH:2003xhc}
H.~Dutz, et~al., First measurement of the {G}erasimov-{D}rell-{H}earn sum rule for {H}-1 from 0.7-{GeV} to 1.8-{GeV} at {ELSA}, Phys. Rev. Lett. 91 (2003) 192001.
\newblock \href {https://doi.org/10.1103/PhysRevLett.91.192001} {\path{doi:10.1103/PhysRevLett.91.192001}}.

\bibitem{Dutz:2004zz}
H.~Dutz, et~al., Experimental check of the {G}erasimov-{D}rell-{H}earn sum rule for {H}-1, Phys. Rev. Lett. 93 (2004) 032003.
\newblock \href {https://doi.org/10.1103/PhysRevLett.93.032003} {\path{doi:10.1103/PhysRevLett.93.032003}}.

\bibitem{Ahrens:2006yx}
J.~Ahrens, et~al., Measurement of the {G}erasimov-{D}rell-{H}earn integrand for {H}-2 from 200-{MeV} to 800-{MeV}, Phys. Rev. Lett. 97 (2006) 202303.
\newblock \href {https://doi.org/10.1103/PhysRevLett.97.202303} {\path{doi:10.1103/PhysRevLett.97.202303}}.

\bibitem{Ahrens:2009zz}
J.~Ahrens, et~al., {Helicity dependence of the total inclusive cross section on the deuteron}, Phys. Lett. B 672 (2009) 328--332.
\newblock \href {https://doi.org/10.1016/j.physletb.2009.01.061} {\path{doi:10.1016/j.physletb.2009.01.061}}.

\bibitem{Helbing:2006zp}
K.~Helbing, The {G}erasimov-{D}rell-{H}earn sum rule, Prog. Part. Nucl. Phys. 57 (2006) 405--469.
\newblock \href {http://arxiv.org/abs/nucl-ex/0603021} {\path{arXiv:nucl-ex/0603021}}, \href {https://doi.org/10.1016/j.ppnp.2005.09.003} {\path{doi:10.1016/j.ppnp.2005.09.003}}.

\bibitem{AguarBartolome:2013mga}
P.~Aguar~Bartolome, et~al., {First measurement of the helicity dependence of $^{3}$He photoreactions in the $\Delta$(1232) resonance region}, Phys. Lett. B 723 (2013) 71--77.
\newblock \href {https://doi.org/10.1016/j.physletb.2013.04.057} {\path{doi:10.1016/j.physletb.2013.04.057}}.

\bibitem{Pedroni:2026eqj}
P.~Pedroni, et~al., {Measurement of the Gerasimov-Drell-Hearn integrand for proton and deuteron from 200 to 1400 MeV} (4 2026).
\newblock \href {http://arxiv.org/abs/2604.14385} {\path{arXiv:2604.14385}}.

\bibitem{GDH:2005noz}
H.~Dutz, et~al., Measurement of helicity-dependent photoabsorption cross sections on the neutron from 815-{MeV} to 1825-{MeV}, Phys. Rev. Lett. 94 (2005) 162001.
\newblock \href {https://doi.org/10.1103/PhysRevLett.94.162001} {\path{doi:10.1103/PhysRevLett.94.162001}}.

\bibitem{Laskaris:2013ehq}
G.~Laskaris, et~al., {First Measurements of Spin-Dependent Double-Differential Cross Sections and the Gerasimov-Drell-Hearn Integrand from $^3\vec He(\vec {\gamma},n)pp$ at Incident Photon Energies of 12.8 and 14.7 MeV}, Phys. Rev. Lett. 110~(20) (2013) 202501.
\newblock \href {http://arxiv.org/abs/1304.5442} {\path{arXiv:1304.5442}}, \href {https://doi.org/10.1103/PhysRevLett.110.202501} {\path{doi:10.1103/PhysRevLett.110.202501}}.

\bibitem{Laskaris:2015wma}
G.~Laskaris, et~al., {Measurement of the doubly-polarized $\vec{{^3}He}(\vec{\gamma},n)pp$ reaction at 16.5 MeV and its implications for the GDH sum rule}, Phys. Lett. B 750 (2015) 547--551.
\newblock \href {http://arxiv.org/abs/1506.00332} {\path{arXiv:1506.00332}}, \href {https://doi.org/10.1016/j.physletb.2015.09.065} {\path{doi:10.1016/j.physletb.2015.09.065}}.

\bibitem{Laskaris:2020ddr}
G.~Laskaris, et~al., {First measurement of the asymmetry and the Gerasimov-Drell-Hearn integrand from the $^{3}$He({\ensuremath{\gamma}},p)$^{2}$H reaction at an incident photon energy of 29 MeV}, Phys. Rev. C 103~(3) (2021) 034311.
\newblock \href {http://arxiv.org/abs/2010.14055} {\path{arXiv:2010.14055}}, \href {https://doi.org/10.1103/PhysRevC.103.034311} {\path{doi:10.1103/PhysRevC.103.034311}}.

\bibitem{Anselmino:1988hn}
M.~Anselmino, B.~L. Ioffe, E.~Leader, On possible resolutions of the spin crisis in the parton model, Sov. J. Nucl. Phys. 49 (1989) 136.

\bibitem{Ji:1999mr}
X.-D. Ji, J.~Osborne, Generalized sum rules for spin dependent structure functions of the nucleon, J. Phys. G 27 (2001) 127.
\newblock \href {http://arxiv.org/abs/hep-ph/9905410} {\path{arXiv:hep-ph/9905410}}, \href {https://doi.org/10.1088/0954-3899/27/1/308} {\path{doi:10.1088/0954-3899/27/1/308}}.

\bibitem{Drechsel:2000ct}
D.~Drechsel, S.~S. Kamalov, L.~Tiator, The {GDH} sum rule and related integrals, Phys. Rev. D 63 (2001) 114010.
\newblock \href {http://arxiv.org/abs/hep-ph/0008306} {\path{arXiv:hep-ph/0008306}}, \href {https://doi.org/10.1103/PhysRevD.63.114010} {\path{doi:10.1103/PhysRevD.63.114010}}.

\bibitem{Drechsel:2004ki}
D.~Drechsel, L.~Tiator, {The Gerasimov-Drell-Hearn sum rule and the spin structure of the nucleon}, Ann. Rev. Nucl. Part. Sci. 54 (2004) 69--114.
\newblock \href {http://arxiv.org/abs/nucl-th/0406059} {\path{arXiv:nucl-th/0406059}}, \href {https://doi.org/10.1146/annurev.nucl.54.070103.181159} {\path{doi:10.1146/annurev.nucl.54.070103.181159}}.

\bibitem{HERMES:2000apm}
A.~Airapetian, et~al., The ${Q}^2$ dependence of the generalized {G}erasimov-{D}rell-{H}earn integral for the proton, Phys. Lett. B 494 (2000) 1--8.
\newblock \href {http://arxiv.org/abs/hep-ex/0008037} {\path{arXiv:hep-ex/0008037}}, \href {https://doi.org/10.1016/S0370-2693(00)01111-4} {\path{doi:10.1016/S0370-2693(00)01111-4}}.

\bibitem{HERMES:2002gmr}
A.~Airapetian, et~al., The ${Q}^2$ dependence of the generalized {G}erasimov-{D}rell-{H}earn integral for the deuteron, proton and neutron, Eur. Phys. J. C 26 (2003) 527--538.
\newblock \href {http://arxiv.org/abs/hep-ex/0210047} {\path{arXiv:hep-ex/0210047}}, \href {https://doi.org/10.1140/epjc/s2002-01118-x} {\path{doi:10.1140/epjc/s2002-01118-x}}.

\bibitem{HERMES:1998pau}
K.~Ackerstaff, et~al., Determination of the deep inelastic contribution to the generalized {G}erasimov-{D}rell-{H}earn integral for the proton and neutron, Phys. Lett. B 444 (1998) 531--538.
\newblock \href {http://arxiv.org/abs/hep-ex/9809015} {\path{arXiv:hep-ex/9809015}}, \href {https://doi.org/10.1016/S0370-2693(98)01396-3} {\path{doi:10.1016/S0370-2693(98)01396-3}}.

\bibitem{Amarian:2002ar}
M.~Amarian, et~al., {The Q**2 evolution of the generalized Gerasimov-Drell-Hearn integral for the neutron using a He-3 target}, Phys. Rev. Lett. 89 (2002) 242301.
\newblock \href {http://arxiv.org/abs/nucl-ex/0205020} {\path{arXiv:nucl-ex/0205020}}, \href {https://doi.org/10.1103/PhysRevLett.89.242301} {\path{doi:10.1103/PhysRevLett.89.242301}}.

\bibitem{CLAS:2008xos}
Y.~Prok, et~al., Moments of the spin structure functions g$^p_1$ and g$^d_1$ for 0.05 \ensuremath{<} ${Q}^2$ \ensuremath{<} 3.0-{GeV}$^2$, Phys. Lett. B 672 (2009) 12--16.
\newblock \href {http://arxiv.org/abs/0802.2232} {\path{arXiv:0802.2232}}, \href {https://doi.org/10.1016/j.physletb.2008.12.063} {\path{doi:10.1016/j.physletb.2008.12.063}}.

\bibitem{CLAS:2015otq}
N.~Guler, et~al., Precise determination of the deuteron spin structure at low to moderate ${Q}^2$ with {CLAS} and extraction of the neutron contribution, Phys. Rev. C 92~(5) (2015) 055201.
\newblock \href {http://arxiv.org/abs/1505.07877} {\path{arXiv:1505.07877}}, \href {https://doi.org/10.1103/PhysRevC.92.055201} {\path{doi:10.1103/PhysRevC.92.055201}}.

\bibitem{CLAS:2017qga}
R.~Fersch, et~al., Determination of the proton spin structure functions for $0.05 < {Q}^2 < 5$~{GeV}$^2$ using {CLAS}, Phys. Rev. C 96~(6) (2017) 065208.
\newblock \href {http://arxiv.org/abs/1706.10289} {\path{arXiv:1706.10289}}, \href {https://doi.org/10.1103/PhysRevC.96.065208} {\path{doi:10.1103/PhysRevC.96.065208}}.

\bibitem{E97-110:2021mxm}
V.~Sulkosky, et~al., Measurement of the generalized spin polarizabilities of the neutron in the low-${Q}^2$ region, Nature Phys. 17~(6) (2021) 687--692, [Erratum: Nature Phys. 18, (2022)].
\newblock \href {http://arxiv.org/abs/2103.03333} {\path{arXiv:2103.03333}}, \href {https://doi.org/10.1038/s41567-021-01245-9} {\path{doi:10.1038/s41567-021-01245-9}}.

\bibitem{JeffersonLabE97-110:2019fsc}
V.~Sulkosky, et~al., Measurement of the 3{H}e spin-structure functions and of neutron (3{H}e) spin-dependent sum rules at 0.035 \ensuremath{\leq} {Q}2 \ensuremath{\leq} 0.24 {GeV}2, Phys. Lett. B 805 (2020) 135428.
\newblock \href {http://arxiv.org/abs/1908.05709} {\path{arXiv:1908.05709}}, \href {https://doi.org/10.1016/j.physletb.2020.135428} {\path{doi:10.1016/j.physletb.2020.135428}}.

\bibitem{CLAS:2021apd}
X.~Zheng, et~al., Measurement of the proton spin structure at long distances, Nature Phys. 17~(6) (2021) 736--741.
\newblock \href {http://arxiv.org/abs/2102.02658} {\path{arXiv:2102.02658}}, \href {https://doi.org/10.1038/s41567-021-01198-z} {\path{doi:10.1038/s41567-021-01198-z}}.

\bibitem{CLAS:2017ozc}
K.~P. Adhikari, et~al., Measurement of the ${Q}^{2}$ dependence of the deuteron spin structure function ${g}_{1}$ and its moments at low ${Q}^{2}$ with {CLAS}, Phys. Rev. Lett. 120~(6) (2018) 062501.
\newblock \href {http://arxiv.org/abs/1711.01974} {\path{arXiv:1711.01974}}, \href {https://doi.org/10.1103/PhysRevLett.120.062501} {\path{doi:10.1103/PhysRevLett.120.062501}}.

\bibitem{CLAS:2024fcf}
A.~Deur, et~al., {Measurement of the nucleon spin structure functions for 0.01{\ensuremath{<}}Q2{\ensuremath{<}}1GeV2 using CLAS}, Phys. Rev. C 111~(3) (2025) 035202.
\newblock \href {http://arxiv.org/abs/2409.08365} {\path{arXiv:2409.08365}}, \href {https://doi.org/10.1103/PhysRevC.111.035202} {\path{doi:10.1103/PhysRevC.111.035202}}.

\bibitem{Deur:2021klh}
A.~Deur, et~al., Experimental study of the behavior of the {B}jorken sum at very low {Q}2, Phys. Lett. B 825 (2022) 136878.
\newblock \href {http://arxiv.org/abs/2107.08133} {\path{arXiv:2107.08133}}, \href {https://doi.org/10.1016/j.physletb.2022.136878} {\path{doi:10.1016/j.physletb.2022.136878}}.

\bibitem{Bernard:1995dp}
V.~Bernard, N.~Kaiser, U.-G. Meissner, Chiral dynamics in nucleons and nuclei, Int. J. Mod. Phys. E 4 (1995) 193--346.
\newblock \href {http://arxiv.org/abs/hep-ph/9501384} {\path{arXiv:hep-ph/9501384}}, \href {https://doi.org/10.1142/S0218301395000092} {\path{doi:10.1142/S0218301395000092}}.

\bibitem{Kahn:2008nq}
Y.~Kahn, W.~Melnitchouk, S.~A. Kulagin, {New method for extracting neutron structure functions from nuclear data}, Phys. Rev. C 79 (2009) 035205.
\newblock \href {http://arxiv.org/abs/0809.4308} {\path{arXiv:0809.4308}}, \href {https://doi.org/10.1103/PhysRevC.79.035205} {\path{doi:10.1103/PhysRevC.79.035205}}.

\bibitem{Ethier:2013hna}
J.~J. Ethier, W.~Melnitchouk, {Comparative study of nuclear effects in polarized electron scattering from 3He}, Phys. Rev. C 88~(5) (2013) 054001.
\newblock \href {http://arxiv.org/abs/1308.3723} {\path{arXiv:1308.3723}}, \href {https://doi.org/10.1103/PhysRevC.88.054001} {\path{doi:10.1103/PhysRevC.88.054001}}.

\bibitem{Ethier:2014bua}
J.~J. Ethier, N.~Doshi, S.~Malace, W.~Melnitchouk, {Quasielastic electron-deuteron scattering in the weak binding approximation}, Phys. Rev. C 89 (2014) 065203.
\newblock \href {http://arxiv.org/abs/1402.3910} {\path{arXiv:1402.3910}}, \href {https://doi.org/10.1103/PhysRevC.89.065203} {\path{doi:10.1103/PhysRevC.89.065203}}.

\bibitem{Tropiano:2018quk}
A.~J. Tropiano, J.~J. Ethier, W.~Melnitchouk, N.~Sato, {Deep-inelastic and quasielastic electron scattering from $A=3$ nuclei}, Phys. Rev. C 99~(3) (2019) 035201.
\newblock \href {http://arxiv.org/abs/1811.07668} {\path{arXiv:1811.07668}}, \href {https://doi.org/10.1103/PhysRevC.99.035201} {\path{doi:10.1103/PhysRevC.99.035201}}.

\bibitem{CLAS:2003umf}
B.~A. Mecking, et~al., The {CEBAF} {L}arge {A}cceptance {S}pectrometer ({CLAS}), Nucl. Instrum. Meth. A 503 (2003) 513--553.
\newblock \href {https://doi.org/10.1016/S0168-9002(03)01001-5} {\path{doi:10.1016/S0168-9002(03)01001-5}}.

\bibitem{Alcorn:2004sb}
J.~Alcorn, et~al., {Basic Instrumentation for Hall A at Jefferson Lab}, Nucl. Instrum. Meth. A 522 (2004) 294--346.
\newblock \href {https://doi.org/10.1016/j.nima.2003.11.415} {\path{doi:10.1016/j.nima.2003.11.415}}.

\bibitem{Strakovsky:2022tvu}
I.~Strakovsky, S.~\v{S}irca, W.~J. Briscoe, A.~Deur, A.~Schmidt, R.~L. Workman, Single-pion contribution to the {G}erasimov-{D}rell-{H}earn sum rule and related integrals, Phys. Rev. C 105~(4) (2022) 045202.
\newblock \href {http://arxiv.org/abs/2201.06495} {\path{arXiv:2201.06495}}, \href {https://doi.org/10.1103/PhysRevC.105.045202} {\path{doi:10.1103/PhysRevC.105.045202}}.

\bibitem{Fix:2005if}
A.~Fix, H.~Arenhoevel, {Double pion photoproduction on nucleon and deuteron}, Eur. Phys. J. A 25 (2005) 115--135.
\newblock \href {http://arxiv.org/abs/nucl-th/0503042} {\path{arXiv:nucl-th/0503042}}, \href {https://doi.org/10.1140/epja/i2005-10067-5} {\path{doi:10.1140/epja/i2005-10067-5}}.

\bibitem{Tiator:2018heh}
L.~Tiator, M.~Gorchtein, V.~L. Kashevarov, K.~Nikonov, M.~Ostrick, M.~Had{\v{z}}imehmedovi{\'c}, R.~Omerovi{\'c}, H.~Osmanovi{\'c}, J.~Stahov, A.~{\v{S}}varc, {Eta and Etaprime Photoproduction on the Nucleon with the Isobar Model EtaMAID2018}, Eur. Phys. J. A 54~(12) (2018) 210.
\newblock \href {http://arxiv.org/abs/1807.04525} {\path{arXiv:1807.04525}}, \href {https://doi.org/10.1140/epja/i2018-12643-x} {\path{doi:10.1140/epja/i2018-12643-x}}.

\bibitem{Matveev:2019igl}
M.~Matveev, A.~V. Sarantsev, V.~A. Nikonov, A.~V. Anisovich, U.~Thoma, E.~Klempt, {Hyperon I: Partial-wave amplitudes for K$^{-}$p scattering}, Eur. Phys. J. A 55~(10) (2019) 179.
\newblock \href {http://arxiv.org/abs/1907.03645} {\path{arXiv:1907.03645}}, \href {https://doi.org/10.1140/epja/i2019-12878-y} {\path{doi:10.1140/epja/i2019-12878-y}}.

\bibitem{Sarantsev:2019xxm}
A.~V. Sarantsev, M.~Matveev, V.~A. Nikonov, A.~V. Anisovich, U.~Thoma, E.~Klempt, {Hyperon II: Properties of excited hyperons}, Eur. Phys. J. A 55~(10) (2019) 180.
\newblock \href {http://arxiv.org/abs/1907.13387} {\path{arXiv:1907.13387}}, \href {https://doi.org/10.1140/epja/i2019-12880-5} {\path{doi:10.1140/epja/i2019-12880-5}}.

\bibitem{Arenhovel:2004ha}
H.~Arenhovel, A.~Fix, M.~Schwamb, {Spin asymmetry and Gerasimov-Drell-Hearn sum rule for the deuteron}, Phys. Rev. Lett. 93 (2004) 202301.
\newblock \href {http://arxiv.org/abs/nucl-th/0407058} {\path{arXiv:nucl-th/0407058}}, \href {https://doi.org/10.1103/PhysRevLett.93.202301} {\path{doi:10.1103/PhysRevLett.93.202301}}.

\bibitem{Workman:2012jf}
R.~L. Workman, M.~W. Paris, W.~J. Briscoe, I.~I. Strakovsky, {Unified Chew-Mandelstam SAID analysis of pion photoproduction data}, Phys. Rev. C 86 (2012) 015202.
\newblock \href {http://arxiv.org/abs/1202.0845} {\path{arXiv:1202.0845}}, \href {https://doi.org/10.1103/PhysRevC.86.015202} {\path{doi:10.1103/PhysRevC.86.015202}}.

\bibitem{Briscoe:2023gmb}
W.~J. Briscoe, A.~Schmidt, I.~Strakovsky, R.~L. Workman, A.~Svarc, {Extended SAID partial-wave analysis of pion photoproduction}, Phys. Rev. C 108~(6) (2023) 065205.
\newblock \href {http://arxiv.org/abs/2309.06631} {\path{arXiv:2309.06631}}, \href {https://doi.org/10.1103/PhysRevC.108.065205} {\path{doi:10.1103/PhysRevC.108.065205}}.

\bibitem{Burkert:2000qm}
V.~D. Burkert, {Comment on the generalized Gerasimov-Drell-Hearn sum rule in chiral perturbation theory}, Phys. Rev. D 63 (2001) 097904.
\newblock \href {http://arxiv.org/abs/nucl-th/0004001} {\path{arXiv:nucl-th/0004001}}, \href {https://doi.org/10.1103/PhysRevD.63.097904} {\path{doi:10.1103/PhysRevD.63.097904}}.

\bibitem{E01-012:2013fky}
P.~Solvignon, et~al., {Moments of the neutron $g_2$ structure function at intermediate $Q^2$}, Phys. Rev. C 92~(1) (2015) 015208.
\newblock \href {http://arxiv.org/abs/1304.4497} {\path{arXiv:1304.4497}}, \href {https://doi.org/10.1103/PhysRevC.92.015208} {\path{doi:10.1103/PhysRevC.92.015208}}.

\bibitem{JeffersonLabHallA:2016neg}
D.~Flay, et~al., {Measurements of $d_{2}^{n}$ and $A_{1}^{n}$: Probing the neutron spin structure}, Phys. Rev. D 94~(5) (2016) 052003.
\newblock \href {http://arxiv.org/abs/1603.03612} {\path{arXiv:1603.03612}}, \href {https://doi.org/10.1103/PhysRevD.94.052003} {\path{doi:10.1103/PhysRevD.94.052003}}.

\bibitem{Bjorken:1968dy}
J.~D. Bjorken, Asymptotic sum rules at infinite momentum, Phys. Rev. 179 (1969) 1547--1553.
\newblock \href {https://doi.org/10.1103/PhysRev.179.1547} {\path{doi:10.1103/PhysRev.179.1547}}.

\bibitem{Deur:2023dzc}
A.~Deur, S.~J. Brodsky, C.~D. Roberts, {QCD} running couplings and effective charges, Prog. Part. Nucl. Phys. 134 (2024) 104081.
\newblock \href {http://arxiv.org/abs/2303.00723} {\path{arXiv:2303.00723}}, \href {https://doi.org/10.1016/j.ppnp.2023.104081} {\path{doi:10.1016/j.ppnp.2023.104081}}.

\bibitem{Brodsky:2008be}
S.~J. Brodsky, R.~Shrock, {Maximum Wavelength of Confined Quarks and Gluons and Properties of Quantum Chromodynamics}, Phys. Lett. B 666 (2008) 95--99.
\newblock \href {http://arxiv.org/abs/0806.1535} {\path{arXiv:0806.1535}}, \href {https://doi.org/10.1016/j.physletb.2008.06.054} {\path{doi:10.1016/j.physletb.2008.06.054}}.

\end{thebibliography}

\end{document}